\documentclass[useAMS,usenatbib]{mn2e}
\usepackage{natbib}
\usepackage{graphicx}
\usepackage{amsmath}
\usepackage{amssymb,latexsym,amsopn}
{\tiny }\usepackage{epsfig}
\usepackage[T1]{fontenc}

\usepackage{color}

\usepackage{ulem}

\title[Neutron star progenitors for FRBs?]{A neutron star progenitor for FRBs? Insights from polarisation measurements}

\author[Ravi \& Lasky]{Vikram Ravi$^{1,}$\thanks{E-mail: vikram@caltech.edu} and Paul D. Lasky$^{2}$
\\
$^{1}$Cahill Center for Astronomy and Astrophysics, MC 249-17, California Institute of Technology, Pasadena, CA 91125, USA.\\
$^{2}$Monash Centre for Astrophysics, School of Physics and Astronomy, Monash University, PO Box 27, VIC 3800, Australia.\\
}

\date{\today}

\pagerange{\pageref{firstpage}--\pageref{lastpage}} \pubyear{2015}

\def\LaTeX{L\kern-.36em\raise.3ex\hbox{a}\kern-.15em
    T\kern-.1667em\lower.7ex\hbox{E}\kern-.125emX}

\begin{document}

\label{firstpage}

\maketitle

\begin{abstract}
	
	
Fast Radio Bursts (FRBs) are intense, millisecond-duration broadband radio transients, the emission mechanisms of which are not understood.  
Masui et al. recently presented Green Bank Telescope observations of FRB~110523, which displayed temporal variation of the linear polarisation position angle (PA). This 
effect is commonly seen in radio pulsars and is attributed to a changing projected magnetic field orientation in the emission region as the star rotates. 
If a neutron star is the progenitor of this FRB, and the emission mechanism is pulsar-like, we show that the progenitor is either rapidly rotating, or the 
emission originates from a region of complex magnetic field geometry. 
The observed PA variation could also be caused by propagation effects within a neutron-star 
magnetosphere, or by spatially varying magnetic fields if the progenitor lies in a dense, highly magnetised environment. Although we urge caution in 
generalising results from FRB~110523 to the broader FRB population, our analysis serves as a guide to interpreting future polarisation measurements of FRBs, and 
presents another means of elucidating the origins of these enigmatic ephemera.
	
\end{abstract}


\section{Introduction}

Over the last 15 years, high-time resolution radio surveys at frequencies between 700\,MHz and 1.5\,GHz have revealed a set of 
millisecond-timescale, broadband transients (fast radio bursts; FRBs) that are possibly extragalactic. The 16 published FRB discoveries are 
summarised by \citet{pbj+16}. The defining characteristics of FRBs are levels of cold-plasma 
dispersion that are unambiguously inconsistent \citep[for discussion, see][]{kon+14} with the Milky Way free-electron density structure, 
and the lack of observed repeat bursts with the same dispersion measure (DM). Poor FRB localisations, limited 
follow-up observations and the lack of credible transient counterparts in the positional error boxes have left the nature(s) of 
FRBs open to untamed speculation. In this paper, assuming that FRBs are extragalactic, we focus on the possibility that FRBs represent extreme 
forms of the radio emission we observe from Milky Way pulsars \citep{cw15} or magnetars \citep{tsb+13,pp10,pp13,kon15}. 

FRB~110523 was discovered at the Green Bank Telescope by \citet{mls+15}, hereafter M15, in the 700\,MHz to 900\,MHz 
frequency band. This burst was 44\% linearly polarised, which enabled the measurement of Faraday rotation of the position angle (PA) of the polarisation 
plane. The degree of Faraday rotation was much larger than expected from the specific Milky Way sightline, and from the intergalactic medium. 
This, combined with evidence for strong scattering from a turbulent medium close to the progenitor, led M15 to infer that the burst 
propagated through a dense, magnetised plasma in its host system. Remarkably, after correcting for the measured 
Faraday rotation, M15 found a monotonically varying polarisation PA of $\sim20^{\circ}$ within the 2\,ms temporal evolution of the FRB, 
which is difficult to attribute to instrumental effects. 

The PA variation in FRB~110523 is tantalisingly reminiscent of the pulses of radio pulsars. There, the linear polarisation vector  
is assumed to be aligned with the magnetic field orientation in the emission region. Hence, linearly polarised radiation from pulsars is used to infer the 
orientation of the rotation and magnetic axes \citep[e.g.,][]{ew01}.  Assuming a global dipole field structure leads to a straightforward relation between the 
polarisation PA, the obliquity angle between the magnetic and rotation axes, $\alpha$, and the smallest inclination angle, $\beta$, between the line of 
sight and the magnetic pole associated with the emission. This relation, known as the rotating vector model \citep[RVM;][]{rc69}, is
\begin{equation}
\tan({\rm PA} - {\rm PA_{0}}) = \frac{\sin(\phi-\phi_{0})\sin\alpha}{\sin\zeta\cos\alpha-\cos\zeta\sin\alpha\cos(\phi-\phi_{0})},
\end{equation}
where $\phi$ is the pulsar rotation phase, $\zeta=\alpha+\beta$, and ${\rm PA_{0}}$ is the PA at a reference rotation phase $\phi_{0}$. 
The remarkable success of the RVM for the Vela pulsar provided important early evidence in support of the 
rotating magnetised neutron star model of pulsars \citep{rck+69}. Measurements of $\alpha$ for large pulsar samples using 
RVM fits revealed that magnetic poles migrate towards the rotation equators as pulsars age \citep{cb86,lm88}. Additionally, measurements of $\beta$ 
have been used to derive pulsar emission beam geometries and solid angles \citep[e.g.,][]{nv82,lm88,tm98}.  

Here, we pursue the idea that the PA variation exhibited by FRB~110523 represents a varying magnetic field orientation within the emission region, 
and derive constraints on the progenitor of the burst assuming the RVM. Although we remain agnostic about whether all reported FRBs 
have similar origins, our constraints can be generalised to all FRBs if such an assumption is made. In \S2, we analyse the effects of the inferred 
multi-path propagation of the burst on the polarisation properties, 
and show that the intrinsic PA variation is likely to have been twice as large as was observed. Then, in \S3, we constrain the orientation of the 
magnetic and rotation axes for different 
possible rotation periods of a neutron-star progenitor of the burst. We discuss the implications of, and various caveats to, our results in \S4, and conclude 
in \S5.

\section{Effects of propagation on the polarisation position angle}

\subsection{Plasma dispersion and scattering}

\begin{figure*}
\centering
\includegraphics[scale=0.9,angle=-90]{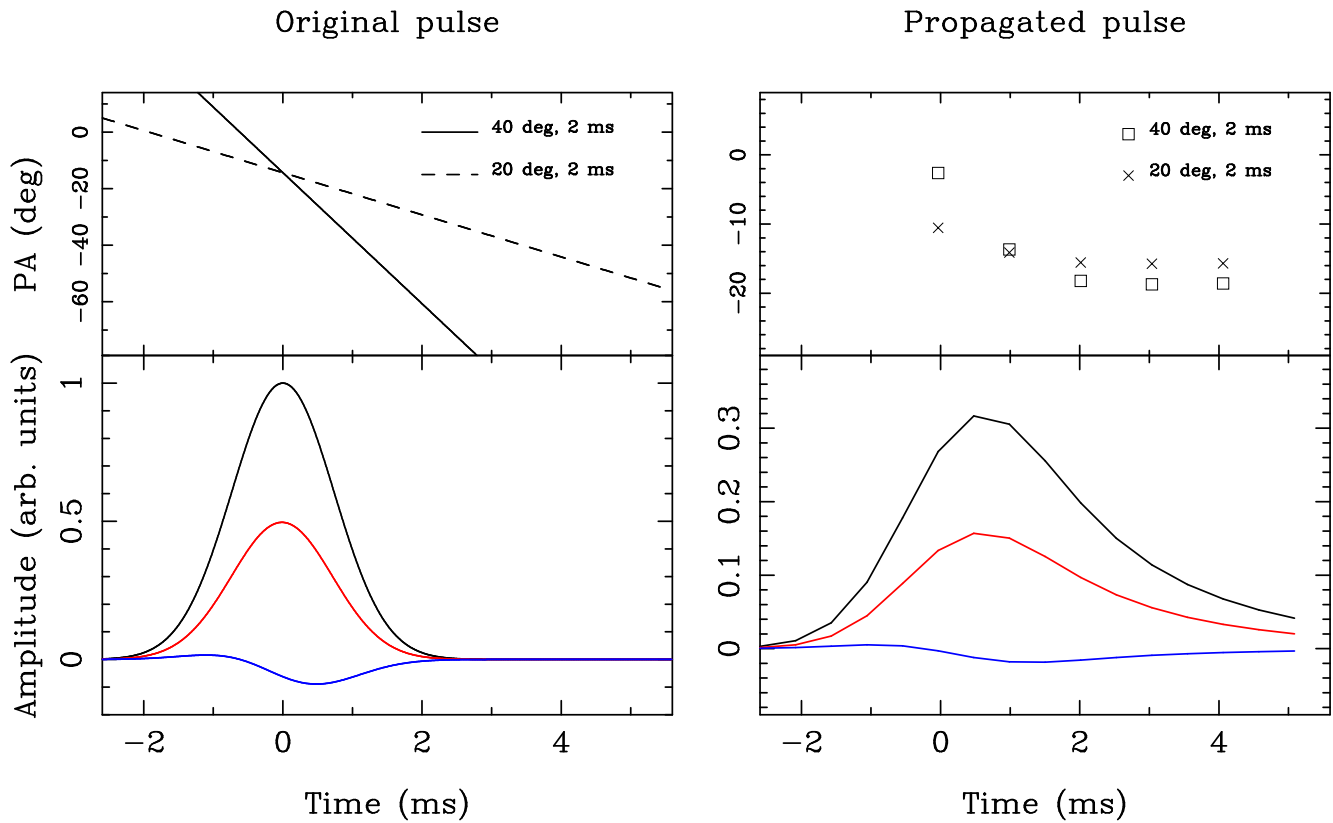}
\caption{Demonstration of the effects of scattering and incoherent dedispersion on FRB~110523. The original pulse 
profile is shown on the left, and the scattered, incoherently dedispersed profile is shown on the right. We have exactly simulated the receiving system 
described by M15 to model the propagated pulse profile. In the bottom 
panels, the black curves represent Stokes I, the red curves represent Stokes Q and the blue curves represent Stokes U. In the top panels, 
the PAs corresponding to the pulse profiles in the bottom panels are shown as a solid line and open squares. The dashed line and crosses 
show a case where an intrinsic PA variation of 20$^{\circ}$ over 2\,ms is assumed; the original polarisation PA variation  
was chosen such that, upon propagation, the results approximately match the data shown in Figure 2 of M15.}
\end{figure*} 

Propagation through tenuous, cold plasma causes radiation to be dispersed, with a group delay given by 
\begin{equation}
t_{d}(\nu) = 4.15 \times \frac{\rm DM}{\rm cm^{-3}\,pc}\left(\frac{\nu}{\rm 1\,GHz}\right)^{-2}\,{\rm ms},
\end{equation}
where $\nu$ is the radio frequency. 
For radio telescope receiver and spectrometer systems with sufficiently large bandwidths to detect FRBs, which have have been found with 
${\rm DM}\gtrsim 100$\,cm$^{-3}$\,pc,
the dispersion delay across these bandwidths is of order a second and must be corrected for. This is achieved by recording data in 
numerous small frequency channels, and delaying the data according to Equation~(2) to align the channels in time for a given DM. This technique, known as 
incoherent dedispersion, has the disadvantage of not correcting for dispersion within each frequency channel: the 
temporal smearing in a single channel of bandwidth $\Delta\nu$ is given by 
\begin{equation}
\Delta t_{d} = 8.3\times \frac{\rm DM}{\rm cm^{-3}\,pc}\left(\frac{\Delta\nu}{\rm 1 GHz}\right)\left(\frac{\nu}{\rm 1\,GHz}\right)^{-3}\,{\rm ms}.
\end{equation}

Additionally, GHz radiation from compact sources is often observed to be scattered by free-electron density inhomegeneities \citep{barney90}. 
Scattering results in a single impulse propagating over multiple paths to the observer, with delays, $t$, relative to the shortest path that are 
typically distributed proportionally to \citep{w72}
\begin{equation}
g(t) = e^{-t/\tau_{s}}.
\end{equation}
Here, $\tau_{s}$ is the characteristic propagation delay caused by the scattering, which is frequency dependent (that is, 
$\tau_{s}\propto \nu^{-\alpha}$). For a scattering medium that is well-modelled by a thin screen, $\alpha=4$ 
for a normal distribution of density inhomogeneities and $\alpha=4.4$ for a Kolmogorov distribution. 

Evidence for multi-path propagation caused by scattering has been observed for eleven FRBs, including FRB~110523. In a few cases 
\citep[M15;][]{tsb+13,rsj15}, the FRB pulse profile has been best modelled by the pulse-broadening function in Equation~(4) convolved with 
narrower profiles. In other events, the pulse width was found to strongly increase with decreasing frequency.  For FRB~110523, the effects of 
both pulse broadening and frequency scintillation were observed. A characteristic scattering propagation delay of $\tau_{s}$ implies a 
scintillation bandwidth of $\sim(2\pi\tau_{s})^{-1}$. Hence, the observation of a scatter broadening timescale of 1.66\,ms at 800\,MHz, combined 
with a scintillation decorrelation bandwidth of 1.2\,MHz was interpreted by M15 as suggestive of two instances of scattering, each 
imparting different characteristic propagation delays.

The effects of scatter broadening on the polarisation properties of pulsar pulse profiles have been investigated by \citet{lh03} and \citet{k09}. 
Unlike scintillation, the observation of pulse broadening caused by scattering implies that each time-frequency element of the pulse corresponds to the 
incoherent sum of many different scattered rays, with different propagation delays. Each scattered ray preserves its intrinsic polarisation properties. 
However, because scatter broadening causes signals corresponding to different parts of the intrinsic pulse to be mixed into each 
time-frequency element, a pulse that has intrinsically varying polarisation properties will be (\textit{a}) depolarised and (\textit{b}) appear to have a 
less-variable polarisation PA. These effects are exacerbated by dispersion smearing caused by incoherent dedispersion. 

We demonstrate these effects in Figure~1, where we attempt to match the observed polarisation PA measurements for FRB~110523 to the intrinsic emission 
properties given the effects of scatter broadening and incoherent dedispersion. The left panel shows our adopted intrinsic pulse, and the right panel 
shows the simulated measurements of this pulse, after scatter broadening, using the M15 spectrometer system 
(4096 channels between 600\,MHz and 800\,MHz) and sampling time (1.024\,ms). The simulated observations approximately match the PA measurements 
presented in Figure~3 of M15, with the exception of one egregious datum that has large uncertainty. We used the 
scatter broadening timescale and frequency-dependency ($\tau_{s}=1.66$\,ms at 800\,MHz, $\alpha=3.6$), and the intrinsic pulse full-width 
half-maximum of 1.73\,ms, reported by M15. We assumed a Gaussian intrinsic pulse profile, a linear polarisation percentage of 50\% (the exact 
percentage is irrelevant to the analysis), 
and a linear intrinsic PA variation across the pulse.\footnote{M15 in fact attempted a statistical fit of a linear intrinsic PA variation to their data, 
with limited success. This was attributed to substructure in the intrinsic pulse beyond the linear model.} 
The adopted model uniquely corresponds to the simulated observations if the PA does not 
vary by more than $2\pi$ within the sampling time. Although the most rapid PA variation observed by M15 was by $\sim20^{\circ}$ over $\sim2$\,ms, 
we find that the intrinsic pulse PA may have varied by $\sim40^{\circ}$ over this time. We also show a case where an intrinsic PA variation of 
20$^{\circ}$ over 2\,ms is assumed (the dashed line and crosses in the top panels of Figure~1). The propagated PA variation in this case clearly 
does not match the data of M15.

\begin{figure*}
\centering
\includegraphics[scale=0.9,angle=-90]{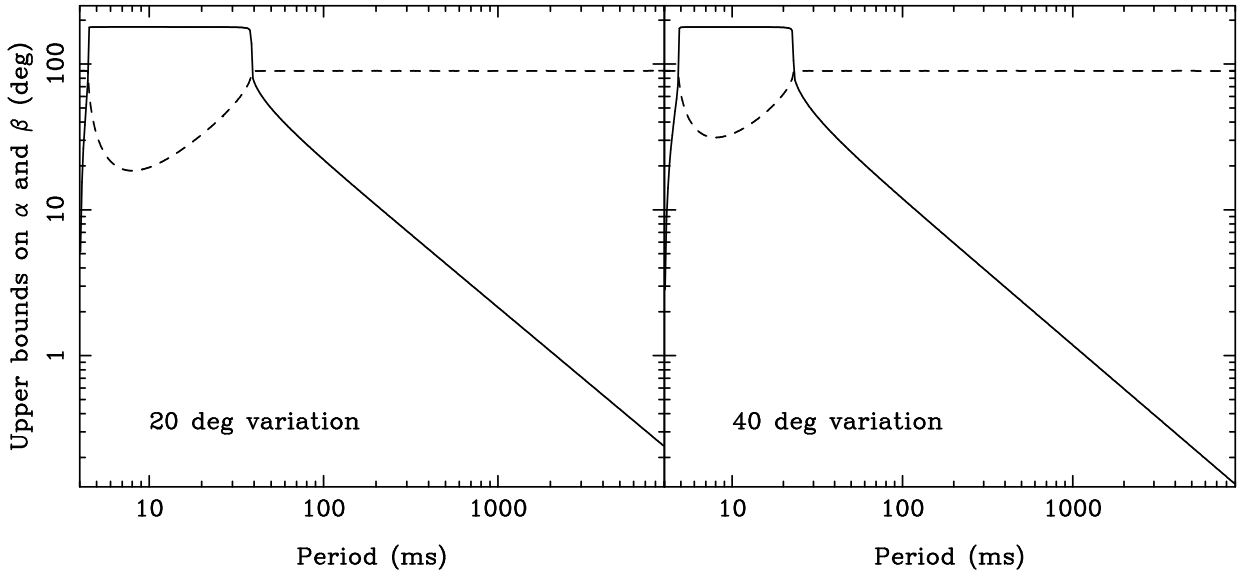}
\caption{Upper bounds on the possible range of magnetic field inclinations, $\alpha$, and impact parameters, $\beta$, 
for a pulsar-like progenitor of FRB~110523. The magnetic field inclination is the angle between the magnetic and rotation axes, and the 
impact parameter is the smallest angle between the line of sight and the magnetic axis. The solid curves show 
upper bounds on $\beta$, and the dashed curves show upper bounds on $\alpha$, for different pulsar periods. On the left, 
we show results for an assumed polarisation position angle (PA) variation of 20$^{\circ}$ in 2\,ms, and on the right we show results 
for a PA variation of 40$^{\circ}$ in 2\,ms. The former is due to the interpretation of M15, whereas the latter 
is the expected true PA variation after accounting for scattering and incoherent dedispersion of the pulse.}
\end{figure*}

\subsection{Path-dependent Faraday rotation}

A further modification of the intrinsic polarisation properties of FRB~110523 could be caused by different amounts of Faraday rotation 
along the different propagation paths implied by the scattering. In this case, both the mixing and the delayed observation of rays that have undergone 
different amounts of Faraday rotation would change the intrinsic PA of the burst. M15 inferred the presence 
of a dense magnetised circum-burst medium, which contributes significantly to the DM of the burst and dominates the Faraday rotation measure (RM). 
The total Faraday rotation over a distance interval between $d_{1}$ to $d_{2}$ is 
\begin{eqnarray}
{\rm RM} & = & 2.6\times10^{-9} \int_{d_{1}}^{d_{2}}n_{e}(s)B_{\parallel}(s)ds \\
& = & 80{\rm DM}_{12}\bar{B}_{\parallel}\,{\rm rad\,m}^{-2}
\end{eqnarray}
where $n_{e}$ is the electron number density, $B_{\parallel}$ is the parallel component of the magnetic field (expressed 
in units of Gauss), $ {\rm DM}_{12}$ is the dispersion measure between $d_{1}$ and $d_{2}$ in the units of cm$^{-3}$\,pc, and $\bar{B}_{\parallel}$ is the 
mean magnetic field over this interval, where the averaging is after weighting by the local electron density, $n_{e}(s)$. For the frequency band in which FRB~110523 was 
observed, two rays propagating over different paths $A$ and $B$ with DM-magnetic field products $({\rm DM}_{12}\bar{B}_{\parallel})_{A}$ and 
$({\rm DM}_{12}\bar{B}_{\parallel})_{B}$ respectively would have a frequency-averaged polarisation PA difference of 
\begin{equation}
\Delta{\rm PA} = 170\times|({\rm DM}_{12}\bar{B}_{\parallel})_{A}-({\rm DM}_{12}\bar{B}_{\parallel})_{B}|\,{\rm deg}.
\end{equation}
Now, the 1.6\,ms scattering time reported by M15 implies a characteristic angular difference between rays of $36[D_{\rm scr}/(1\,{\rm kpc})]^{-1}$\,mas, 
where $D_{\rm scr}$ is the distance of the scattering screen from the source. M15 suggest that $D_{\rm scr}<44$\,kpc based on the 
scattered image being unresolved by a Milky Way scattering screen. 

Let us assume, as M15 conclude, that some of the DM and most of the RM arises close to the burst progenitor. 
Then, to produce the observed PA variation of $\sim20^{\circ}$ over approximately 
the scattering timescale, we require that $|({\rm DM}_{12}\bar{B}_{\parallel})_{A}-({\rm DM}_{12}\bar{B}_{\parallel})_{B}|\sim0.12$\,cm$^{-3}$\,pc\,G. 
For typical interstellar magnetic fields of order microGauss, or even for the extreme milliGauss magnetic fields found in dense star-forming regions, 
this requires unrealistic DM differences between the propagation paths of $1.2\times10^{2}$\,cm$^{-3}$\,pc to  $1.2\times10^{5}$\,cm$^{-3}$\,pc. 
However, if we assign a DM of 100\,cm$^{-3}$\,pc to the circum-burst medium, a parallel magnetic field variation between the propagation paths of only 1.2\,mG is 
required to explain the PA variation across the burst. We stress that these fields are averaged following weighting by the electron density along the path. 
Nonetheless, such a scenario is not entirely unrealistic for a dense star-forming region such as the Orion nebula \citep{kon15}, or for a 
galactic-centre environment \citep[e.g.,][]{efk+13}. We therefore proceed 
to consider the polarisation PA variation of FRB~110523 in the context of the RVM, but with the caveat that the possible origin of the burst 
in a dense star-forming region or galactic-centre environment could also cause the observed PA variation.

\section{The rotating vector model: neutron star constraints}

In this section, we apply the RVM to calculate the maximum allowed ranges of the obliquity, $\alpha$, and the magnetic inclination, $\beta$, of the 
burst progenitor for different rotation periods. Even if we assume that the RVM applies to FRB~110523, the rotation period is coupled to the 
values of $\alpha$ and $\beta$. Qualitatively, to produce the observed PA variation, tighter upper bounds on 
$\beta$ are expected for larger rotation periods, whereas the range of $\alpha$ remains relatively unconstrained. For these constraints, the exact pulse phase 
range assumed for the PA variation is not important.

We consider two cases for the PA variation: one that is due to the interpretation of M15 (20$^{\circ}$ in 2\,ms), and 
one that corresponds to our expectation from \S2 for the intrinsic PA variation before scattering (40$^{\circ}$ in 2\,ms). We show our results in Figure~2 for 
each case. The maximum allowed ranges displayed for $\alpha$ and $\beta$ are calculated from Equation~(1); values outside these ranges 
for each rotation period are not geometrically possible. Rotation periods shorter than 4\,ms are forbidden by the RVM.

We do not attempt to present probabilistic constraints on the angles because our 
lack of prior knowledge of the system makes this of negligible value. We note that for each specific value of the spin period, there is a 
one-to-one mapping between $\alpha$ and $\beta$. For periods between $\sim5$\,ms and $\sim14$\,ms, $\alpha$ is 
mildly constrained because values of $\alpha$ close to $90^{\circ}$ 
would imply too rapid a PA variation for any $\beta$. For larger rotation periods, as expected, the upper bound on $\beta$ decreases. For rotation 
periods $>1$\,s, $\beta\lesssim2^{\circ}$ or $\beta\lesssim1^{\circ}$ depending on the assumed PA variation.

\section{Discussion}

\subsection{Constraints on neutron-star progenitor models}

If FRB~110523 has a neutron-star progenitor, the polarisation profile is consistent with the RVM for a wide range of rotation periods. Although our analysis 
is rudimentary, being based only on the total PA variation over 2\,ms, the constraints we obtain on $\beta$ 
are interesting. However, we emphasise that we cannot distinguish between different progenitor periods, because in our analysis all periods have 
combinations of $\alpha$ and $\beta$ that reproduce the data equally well. We are not bold enough to statistically fit the two-parameter RVM to the three 
highly significant PA measurements for FRB~110523, and this precludes us from undertaking model selection on the progenitor rotation period. 

A number of models for FRB progenitors posit magnetospheric emission from rotating neutron stars. \citet{cw15} suggest that 
an extreme form of giant pulse emission from young pulsars at modest distances of a few hundred Mpc could explain the observed FRB population. 
Matching FRB rates to classes of astrophysical explosions has led multiple authors to associate FRBs with magnetar 
hyper-flares \citep{tsb+13,pp10,pp13,kon15}. However, existing magnetar models for FRBs \citep{l14,katz15} do not naturally predict a polarisation 
PA swing across the pulse in any circumstance. This is because the magnetic pulses expected during hyper-flares from magnetars are predicted to result in 
radio bursts upon interaction with magnetar wind nebulae, which do not co-rotate with the star. On the other hand, 
if we remain agnostic about the actual FRB emission mechanism associated with magnetar hyper-flares, then another option is that emission comes 
from near the surface of the star and is associated with the fireball \citep{td95}. As the rotation periods of known magnetars are all $\gtrsim 1$\,s \citep{ok14}, 
our results may provide stringent constraints on the magnetic field geometry associated with FRBs from hyper-flares, as explained below.

An arbitrarily chosen pulsar or magnetar is more likely to be slow-spinning than fast-spinning, owing to magnetic spin-down. For example, a rotating star with a 
dipole field strength at the pole of $B_{p}$ has a rotation period that evolves with time, $t$, proportionally to $(1+KB_{p}t)^{1/2}$, where $K$ is a constant 
\citep{st86}. 
The majority of radio pulsars have rotation periods $\gtrsim0.1$\,s \citep{mht+05}, and all Galactic magnetars, which have estimated ages of tens of kyr, have even 
larger rotation periods \citep{ok14}. Our constraints on $\alpha$ and $\beta$ indicate that, if FRB~110523 represents pulsar-like emission that 
can be modelled with the RVM, one of two possibilities is allowed:
\begin{enumerate}
\item the burst originates from a fast-rotating (and hence likely young) neutron star, in which case we cannot constrain the emission geometry, or
\item the burst originates from a more slowly rotating neutron star, consistent with the bulk of the pulsar population and the entire observed magnetar population. In 
this case, if the rotation period of the progenitor is $>0.1$\,s,  we constrain the emission to originate from within $20^{\circ}$ of the magnetic pole, and likely within 
$10^{\circ}$ after accounting for the effects of scattering. It is also possible that the emission originates from a magnetic field structure that is more complex, 
and hence possesses greater curvature, than the global dipole field. Such structures are expected as a result of the magnetohydrodynamical 
instabilities that may trigger magnetar hyper-flares \citep[e.g.,][]{td95,lzk+11}. 
\end{enumerate}

Further inferences can be drawn by considering our results in the context of RVM analyses of the emission geometries of pulsars and magnetars. 
Radio emission from fast-rotating pulsars with large spin-down luminosities, including young and millisecond pulsars, is expected to cover large fractions of 
$4\pi$ steradians, and to originate from high in the magnetosphere \citep{rmh10}. Older, more slowly rotating pulsars are generally found to have narrower 
emission beams tied to one or both magnetic poles \citep{lm88,r90,tm98}. Our constraints on $\beta$ for different rotation periods are consistent with, but cannot 
distinguish between, both scenarios. However, it is also found that pulsars are born with large values of $\alpha$ that exhibit secular decreases with age 
\citep{lm88}; this effect has indeed been observed in `real time' for the Crab pulsar \citep{lgw+13}. Hence, if a neutron star progenitor of FRB~110523 
were to rotate slowly, it is possible that it is an aligned rotator, with small values of both $\alpha$ and $\beta$. 

\subsection{Possible contaminating effects}

Although we only have one burst of emission from the progenitor of FRB~110523, this does not imply a stochastic bias in our analysis. On the contrary, 
our single-pulse analysis is less susceptible to contamination by the orthogonally polarised modes (OPMs) commonly observed in pulsars \citep{br80,nv82}. 
Single pulses from many pulsars are observed to switch between two OPMs, which have varying predominance at different pulse longitudes. Such behaviour 
is also observed in rotating radio transients \citep[e.g.,][]{khv+09}. Some care is required to account for OPMs in RVM fits to integrated 
pulse profiles \citep[e.g.,][]{nv82,ew01}. However, our analysis of a single pulse is indeed more optimal than analyses of integrated pulse profiles as the PA 
measurements are not the averages of pulses in different polarisation modes. 

However, even if the progenitor of FRB~110523 is a neutron star, other effects may contaminate our RVM inferences. We have already noted the possibility 
that if the burst originates in a dense, highly magnetised environment, the observed polarisation PA variation could be induced by varying magnetic fields 
along multiple propagation paths (\S2.2). Further, the emission at different times likely 
traverses different paths through the magnetosphere which may result in different degrees of Faraday rotation \citep{km98}. 
If the emission originates at $\lesssim20$\,km from the neutron-star surface, Lense-Thirring precession results in additional observed curvature of 
magnetic field lines \citep{mt92}, and 
aberration and retardation can also cause distortions in the observed field-line geometry for polar-cap emission \citep[e.g.,][]{gg01}. 
Propagation effects in the magnetosphere could also play a role in mimicking PA swings \citep{cr79}. 

Perhaps of more relevance are 
the polarisation properties of giant pulses from pulsars, and individual radio pulses from magnetars. \citet{jpk+10} reported on the polarisation 
properties of nanosecond-timescale giant pulses from the Crab pulsar, which have been considered as direct analogues of FRBs by \citet{cw15}. Although substantial pulse-to-pulse 
variation is seen, some giant pulses show tens of degree PA variations on timescales of a few nanoseconds. Such extreme PA swings are either indicative of 
emission from incredibly close to the magnetic pole, or of the relativistic effects mentioned above or proper motion of the emitting structures. 
Furthermore, evidence of magnetospheric 
propagation-related mixing of linear and circular polarisation is observed in some magnetars \citep{ksj+07}. 

\subsection{Generalising our results to the FRB population}

Clearly, caution is required in generalising our results to the full FRB population. In analogy with the different gamma-ray burst populations observed by 
instruments with different sensitivities and spectral responses, different radio telescopes may be sensitive to different 
populations of FRBs. Besides the Parkes telescope, FRB candidates have only been detected at the Arecibo Observatory (FRB~121102) and the Green Bank 
telescope (FRB~110523). However, we note that the lower limit on the fluence of FRB~110523 reported by M15 of 1.9\,Jy\,ms is comparable 
to the Parkes and Arecibo FRBs. Further, the fluence of FRB~110523 will be attenuated by free-free absorption in the circum-burst medium. The free-free 
optical depth, assuming a temperature of 8000\,K for the circum-burst medium, is \citep[][and references therein]{kon15}
\begin{equation}
\tau_{\rm ff} = 0.47 \frac{\rm DM}{100\,{\rm cm}^{-3}\,{\rm pc}}\frac{\bar{n}_{e}}{5000\,{\rm cm}^{-3}}\left[\frac{\nu(1+z)}{700\,{\rm MHz}}\right],
\end{equation}
where $\bar{n}_{e}$ is the mean electron density along the line of sight, and $z$ is the cosmological redshift. This implies an attenuation factor of 
$\sim1.6$. 

The PA variation exhibited by FRB~110523 implies that the burst is temporally resolved even after accounting for 
scatter broadening. This is consistent with the unscattered pulse full-width half-maximum  of 1.7\,ms reported by M15, which is 
significantly greater than the maximum dispersion smearing timescale within the frequency band (Equation~3) of $\Delta t_{d} = 0.74$\,ms. 
In contrast, neither of the Parkes FRBs for which a pulse broadening function (Equation~4) was fit \citep[FRB~110220, FRB~131104;][]{tsb+13,rsj15} 
were temporally resolved after the scattering was accounted for. For FRB~110220, the upper limit on the intrinsic width set by dispersion smearing was 
1\,ms, and the corresponding upper limit for FRB~131104 was 0.5\,ms. This may indicate a difference between FRB~110523 and the Parkes FRB sample. 

Besides FRB~110523, the only reported polarisation measurement for an FRB (140514) is from Parkes \citep{pbb+15}, where 
the burst was found to be 21\% circularly polarised with no linear polarisation. Although nanosecond-timescale structures in giant pulses from the Crab pulsar 
can be highly circularly polarised, the even split between left and right circular polarisation means that the sum of these structures usually shows only 
linear polarisation \citep{jpk+10}. Circular polarisation in pulsars is also commonly associated with magnetospheric propagation effects \citep{cr79}. Further,  
linear polarisation can be depolarised if Faraday rotation causes rapid wrapping of the phase between Stokes Q and U over individual spectrometer channels; 
however, an RM in excess of $10^{5}$\,rad\,m$^{-2}$ is required for FRB~140514 to be completely depolarised \citep{pbb+15}. Hence, 
as a minimum, further polarisation measurements are required for FRBs from multiple telescopes in order to understand whether the implications 
of our RVM analysis can be generalised to all FRBs. Coherent dedispersion will also likely enable intrinsic pulse widths to be resolved. 
If FRBs represent pulsar-like bursts from rotating, magnetised neutron stars, it is likely that at least some will repeat on observable timescales \citep{cw15}.

\section{Conclusions}

We have examined the variation in the polarisation PA within FRB~110523 in the context of a rotating, magnetised neutron star 
progenitor for the burst. We account for the effects of the observed scattering of the burst by considering two cases for the PA variation: 
$20^{\circ}$ in 2\,ms as reported by M15, and $40^{\circ}$ in 2\,ms, which we demonstrate to approximately correspond to the intrinsic PA variation before scattering. 
By analysing these PA variations with the rotating vector model commonly used to ascertain pulsar emission geometries \citep{rc69}, we find that 
if a neutron-star progenitor of FRB~110523 were slowly-rotating (e.g., with rotation period $>0.1$\,s), the emission must originate from close to 
a magnetic pole (magnetic inclination $\beta<20^{\circ}$, and $<10^{\circ}$ after accounting for scattering). We conclude that
\begin{enumerate}
\item FRB~110523 is consistent with the RVM, given empirical results for pulsars in the Milky Way.
\item If the burst represents pulsar-like emission from a rapidly rotating (and hence likely young) neutron star, we cannot constrain the emission geometry. 
\item If the burst represents pulsar-like emission from a more slowly rotating (and hence likely older) neutron star, the emission may either originate 
from close to a magnetic pole, or from a complex local field structure. In the former case, given the typical alignment between the rotation and 
magnetic axes of older neutron stars, the progenitor may be an aligned rotator. 
\end{enumerate}

However, even if the progenitor of FRB~110523 is a rotating, magnetised neutron star, various effects may bias our RVM analysis. For example, the 
observed PA variation could be caused by (\textit{a}) varying magnetic fields along different propagation paths if the burst originates in a dense, highly 
magnetised region, or (\textit{b}) magnetospheric propagation effects or intrinsic motion of the emitting region. Additionally, we also cannot ascertain whether our 
conclusions apply to some or all FRBs. That is, it is unclear whether all FRBs have the same origin and whether that origin is pulsar-like. To aid in generalising 
our conclusions, further high-quality polarisation measurements with coherent dedispersion instruments are required from multiple telescopes. 

\section*{Acknowledgements}

VR acknowledges enlightening discussions with S. R. Kulkarni, H. Vedantham, E. G. Thomas, R. M. Shannon and J. Sievers. PDL is supported by an 
Australian Research Council Discovery Project DP1410102578.

\bibliographystyle{mn2e}
\bibliography{mn-jour.bib,vikram.bib}

\label{lastpage}

\end{document}